\documentclass[twocolumn, 12pt]{extarticle} 

\usepackage[english]{babel}
\usepackage[utf8]{inputenc}
\usepackage[T1]{fontenc}
\usepackage{siunitx}
\DeclareSIUnit\micron{\micro\metre}

\usepackage{textcomp} 

\usepackage[margin=0.7in]{geometry}
\usepackage{amsmath}
\usepackage{amssymb}
\usepackage{textcomp} 
\usepackage{graphicx} 
\graphicspath{{./images/}}
\usepackage{float}    
\usepackage[colorlinks=true, allcolors=blue]{hyperref}
\usepackage{authblk} 
\usepackage{wrapfig} 

\title{Designing high-performance propagation-compressing spaceplates using thin-film multilayer stacks}
\author[1,2]{Jordan T. R. Pag\'e}
\author[1,*]{Orad Reshef}
\author[1,2,3]{Robert W. Boyd}
\author[1]{Jeff S. Lundeen}
\affil[1]{Department of Physics, University of Ottawa, 25 Templeton Street, Ottawa, ON K1N 6N5, Canada}
\affil[2]{School of Electrical Engineering and Computer Science, University of Ottawa, Ottawa, ON K1N 6N5, Canada}
\affil[3]{Institute of Optics and Department of Physics and Astronomy, University of Rochester, Rochester, NY 14627, USA}
\affil[*]{Corresponding author: orad@reshef.ca}

\date{\vspace{-3em}}

\begin{document}

\twocolumn[
  \begin{@twocolumnfalse}
    \maketitle
    
    \begin{abstract}
      \vspace{0.5em}
      
      \hspace{-1.25cm}\begin{minipage}{0.5\textwidth}
        {\includegraphics[width=3.25in, height=1.75in]{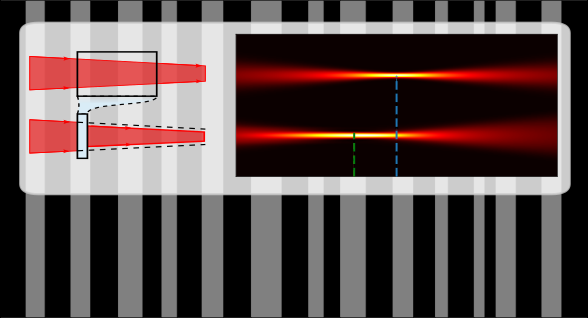}}
      \end{minipage}
      \begin{minipage}{0.5\textwidth}
    \noindent The development of metasurfaces has enabled unprecedented portability and functionality in flat optical devices. Spaceplates have recently been introduced as a complementary element to reduce the space between individual metalenses. This will further miniaturize entire imaging devices.
    However, a spaceplate necessitates a non-local optical response --- one which depends on the transverse spatial frequency component of a light field --- therefore making it challenging both to design them and to assess their ultimate performance and potential. 
    Here, we employ inverse-design techniques to explore the behaviour of general thin-film-based spaceplates. 
    We observe a tradeoff between the compression factor $\mathcal{R}$ and the numerical aperture $\mathrm{NA}$ of such devices; we obtained a compression factor of $\mathcal{R} = 5.5$ for devices with an $\mathrm{NA} = 0.42$ up to a record $\mathcal{R} = 340$ with $\mathrm{NA}$ of 0.017. 
    Our work illustrates that even simple designs consisting of realistic materials (\emph{i.e.,} silicon and glass) permit capable spaceplates for monochromatic applications.
      \end{minipage}
      
    \vspace{1em}
    \end{abstract}
    \noindent\hfil\hspace{1cm}\rule{0.8\textwidth}{.4pt}\hfil
    \vspace{1em}
  \end{@twocolumnfalse}
]

\noindent Nanostructured surfaces known as metasurfaces have been gaining much attention for enabling the creation of flat optics~\cite{Yu2011, Yu2014, Chen2020a}. Among the most notable devices are metalenses, optical elements that promise to yield ultra-thin imaging systems by replacing the relatively thick glass of lenses with a much thinner optic~\cite{Khorasaninejad2016, Khorasaninejad2017, Liang2018}. Unfortunately, these recent advances in miniaturizing lenses do not address the space \emph{between} the numerous optics, sources, and sensors, which make up the largest part of most imaging systems.

\begin{figure}[!tbh]
\begin{centering}
\includegraphics[width=0.6\linewidth]{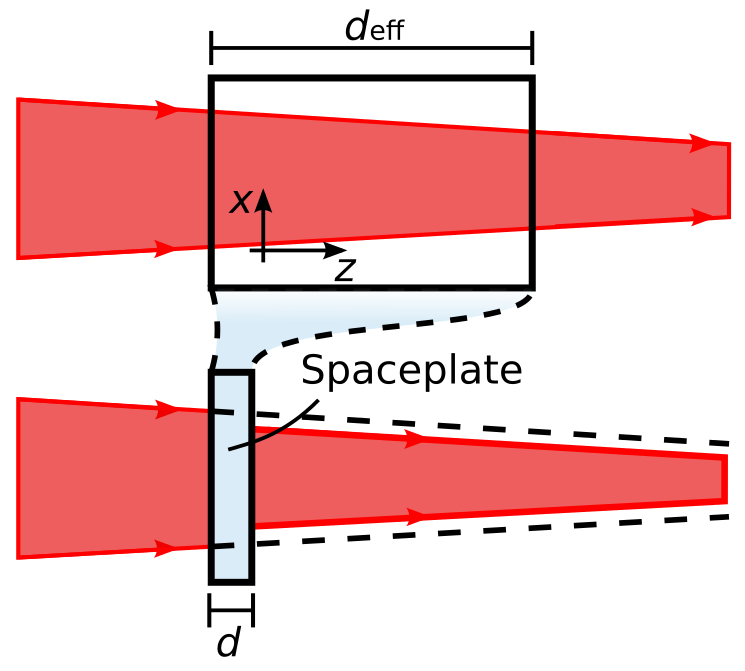}
\par\end{centering}
\caption{A spaceplate compresses the propagation of a length $d_{\mathrm{eff}}$ into a region of length $d$.}
\label{Fig:Spaceplate}
\end{figure}

Recently, a new optical element has been proposed that can simulate free-space propagation to further reduce imaging systems, called a `spaceplate'~\cite{Reshef2021, Guo2020a, Chen2020}. This device consists of a plate of thickness $d$; however a transmitted lightfield will experience an effective propagation of length $d_\mathrm{eff}>d$ (Fig.~\ref{Fig:Spaceplate}). This expanded propagation is accomplished through a non-local operation~\cite{Castaldi2012a, Silva2014, Kwon2018, Cordaro2019} that imparts an angle-dependent phase onto an incident beam. By replacing free-space, spaceplates could one day lead to ultra-thin telescopes, microscopes, and cameras, for example eliminating the camera bump on the back of smartphones.

Past works have experimentally demonstrated spaceplates in the form of a low-index medium or a birefringent crystal~\cite{Reshef2021}, and have theoretically demonstrated spaceplates in the form of thin-film multilayer stacks~\cite{Reshef2021}, a two-dimensional photonic crystal slab~\cite{Guo2020a}, and coupled cavities~\cite{Chen2020}.  
These devices featured compression factors $\mathcal{R}\equiv (d_\mathrm{eff}/d)$ that ranged from $\mathcal{R}=1.1$ to $\mathcal{R}=144$, with other metrics of varying performance, such as numerical aperture (NA), insertion loss, spectral operating bandwidth, and total simulated space $d_\mathrm{eff}$. 
Aside from the special case of resonator-based spaceplates~\cite{Chen2020}, a general design methodology and identification of the fundamental limitations of high-performance spaceplates has remained a challenge. This stands in contrast to other more mature devices, such as metalenses~\cite{Presutti2020a} or invisibility cloaks~\cite{Fleury2015}, for which there are established design prescriptions. As a result, it is as of yet unknown what the fundamental tradeoffs or limitations on these performance metrics may be for this new type of device.

In this Letter, we expand on previous inverse-design work on a class of spaceplates based on multilayer thin-film stacks. Whereas previous work featuring this platform used a genetic algorithm \cite{Reshef2021} and produced a maximum $\mathcal{R}=4.9$, here we develop a gradient descent optimization that produces much higher performance spaceplates. We use this optimization to explore the parameter space for this class of spaceplates and establish limits on their performance. 

A spaceplate will replace a slab of a medium of length $d_\mathrm{eff}$ if the spaceplate acts on every possible incident planewave in the same manner as the slab. In particular, each incident planewave is transmitted without loss but receives an angle-dependent phase of the form 
\begin{equation}
\phi(\theta) = |\mathbf{k}| d_\mathrm{eff} \cos\theta, 
\label{EQ:SpaceplatePhase}
\end{equation}
where  $\theta$ is the incident angle of the light, and $\mathbf{k}$ is the wavevector in the background medium~\cite{Reshef2021}. Throughout this Letter, the background medium is air ($n=1$). Equation~\eqref{EQ:SpaceplatePhase} ensures that a ray exiting a spaceplate will preserve its incident angle just as it would when travelling through air. Equation~\eqref{EQ:SpaceplatePhase} also leads to a translation of the ray's lateral position as illustrated in Fig.~\ref{Fig:Spaceplate}. This angle-conservation is contrary to the behaviour of a ray at a lens, where the ray angle changes due to refraction. In this way, lenses and spaceplates are complementary devices.

In order to quantify the performance of the spaceplates that we generate through inverse design, we use the Strehl ratio~$S$~\cite{Strehl}. This quantity is commonly used to measure the quality of lenses. It is the ratio of the peak intensity of a focus of a Gaussian beam that travels through a real (\emph{i.e.,} aberrated) lens to the peak intensity created by an ideal lens. $S$ ranges 0 to 1. A Strehl ratio of $S \geq 0.8$ is standardly diffraction limited~\cite{Marechal}. Similarly, a real spaceplate will impart phase errors to a beam that is already focusing through it and so the Strehl ratio is also a relevant metric. The Strehl ratio is given by the formula:
\begin{equation}
    S = e^{-\sigma^2}
\end{equation}
where $\sigma$ is the root-mean-square-error (RMSE) between the phase of an ideal spaceplate and a designed spaceplate of the same thickness $d$. We found $\mathcal{R}$ for the latter from a fit to its the transmission phase. A target NA for the designed spaceplate sets the range of angles over which the RMSE is calculated.

In our exploration, we focus on multilayer stacks which can be easily and accurately modeled using the transfer-matrix method (TMM)~\cite{Yariv2007a}.
This method is used to obtain the phase and the transmission coefficients of a stack. 
We consider $p$-polarized light at an operating wavelength of $\lambda=1550$~nm, thus restricting our study to monochromatic operation. 
Monochromatic spaceplates could be used to reduce the size of sensors, monochromatic displays or spatial multiplexing applications, where we only utilize narrow bands of frequencies. 
We use gradient descent as our optimization technique, where we define the figure of merit (FOM) that we maximize to be $1/\sigma$.  
Details of the optimization procedure are given in the Methods section. 
Unlike with the Strehl ratio, the RMSE $\sigma$ is taken between the phase of the candidate spaceplate and an ideal spaceplate with $d_\mathrm{eff}=\mathcal{R}_{\mathrm{target}}d$.  
Each device optimization thus targets a $\mathcal{R}_{\mathrm{target}}$ and NA.
We explore two approaches, one that uses two materials and optimizes the layer thicknesses, and a second that uses fixed layer thicknesses and optimizes each layer's refractive index.

\begin{figure}[!htb]
\begin{centering}
\includegraphics[width=1\linewidth]{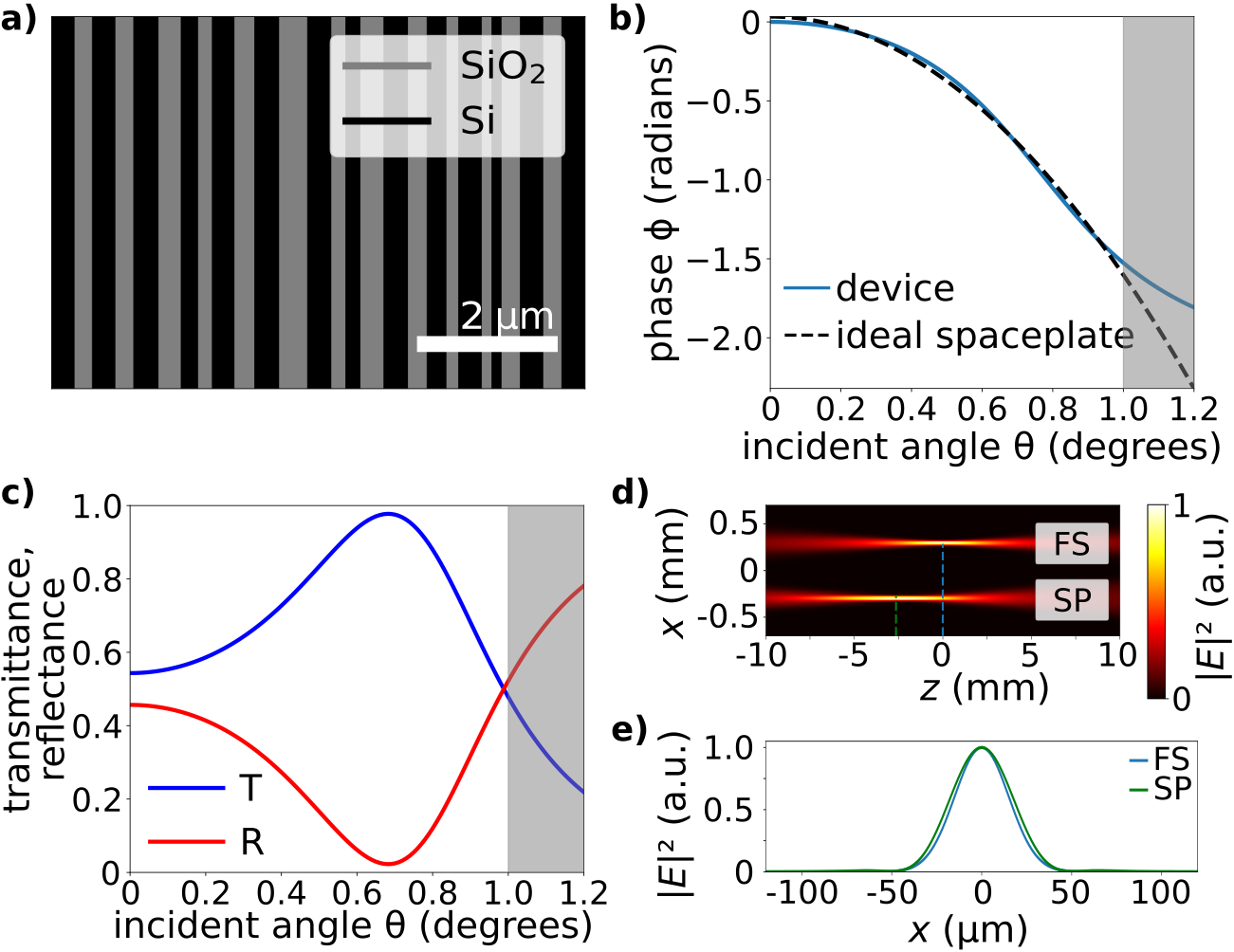}
\par\end{centering}
\caption{{\bf a)}~A spaceplate consisting of 27 alternating layers of Si (black) and SiO$_2$ (grey) with a total thickness of \SI{7.6}{\micron}, which is surrounded by air. The layers are drawn to scale with respect to the scale bar (\SI{2}{\micron}). {\bf b)}~The transmission phase of the device fit to the expected output phase of an ideal spaceplate with an effective advance of $d_\mathrm{eff}=2.6$~mm. It behaves as a spaceplate with a compression factor of $\mathcal{R}=340$ for $\mathrm{NA} = 0.017$ (up to $1$\textdegree{} --- the unshaded region). {\bf c)}~The normalized transmittance and reflectance of the spaceplate. This device has an angle-dependent transmission profile, with a minimum insertion loss of 3.3~dB at an incident angle of 1\textdegree{}. {\bf d)}~Analytic Fourier optics propagation is used to find the intensity $|E|^2$ of a focusing Gaussian beam (waist of \SI{30}{\micron}, divergence of 0.94\textdegree{}) propagating in free space (FP) and after propagating through the metamaterial spaceplate (SP) from left to right. {\bf e)}~Beam cross section along the vertical dashed lines in d). }
\label{Fig:R=340}
\end{figure}

The first approach considers a structure of alternating layers of silicon ($n=3.48$) and silica ($n=1.44$). 
The structure is surrounded on both sides by air ($n=1$). 
Here, we set a fixed number of thin-film layers and the thicknesses of the individual layers are used as the optimization parameters.
Figure~\ref{Fig:R=340} illustrates a resulting device consisting of 27 layers optimized for operation for a maximum half-angle of 1\textdegree{} ($\mathrm{NA}=0.017$). 
The device is \SI{7.6}{\micron} thick and replaces propagation for 2.6~mm, corresponding to a compression factor of $\mathcal{R} = 340$. 
This value is the largest observed in any study.
It has a maximum insertion loss of only 3.3~dB over its designated angular operating bandwidth. 
Figure~\ref{Fig:R=340}b gives the imparted phase of the device (blue line), which fits well to the ideal spaceplate curve (dashed line, given by Eq.~\ref{EQ:SpaceplatePhase}).
This device has a Strehl ratio of $S=0.9992$, far above diffraction-limited performance.

To further probe the potential performance of the thin-film multilayer stack platform as a spaceplate, we explored the parameter space in a thorough series of simulations. Though a direct search of this type is not in any sense exhaustive, its results can be instructive in drawing certain conclusions, as we shall see. In our study, for a given number of layers ranging from 9~--~29, we swept over  targeted compression factors from $\mathcal{R}_\mathrm{target}=2$~--~60 and explored devices with five NAs: 0.087, 0.17, 0.26, 0.34, and 0.42, corresponding to maximum half-angles of incidence of 5\textdegree{}, 10\textdegree{}, 15\textdegree{}, 20\textdegree{} and 25\textdegree{}, respectively. 

\begin{figure}[!b]
\begin{centering}
\includegraphics[width=1\linewidth]{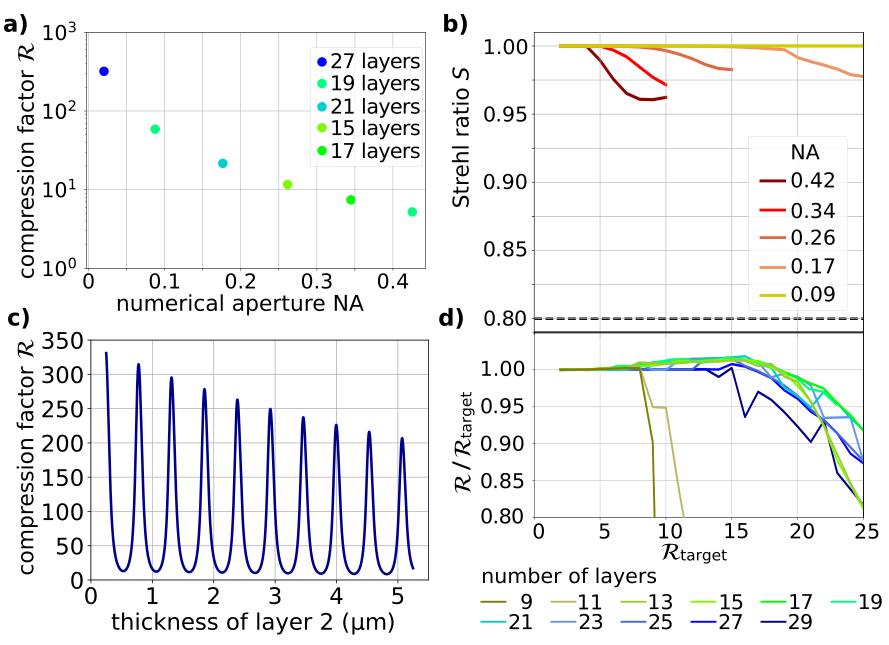}
\par\end{centering}
\caption{{\bf a)}~ The maximum observed compression factor $\mathcal{R}$ as a function of NA of any spaceplate with 9, 11, ..., 27 or 29 layers. 
The compression factor decreases with increasing numerical apertures. 
{\bf b)}~The Strehl ratio $S$ as a function of $\mathcal{R}_{\mathrm{target}}$ for devices with acceptance angles of 5\textdegree{}, 10\textdegree{}, 15\textdegree{}, 20\textdegree{}, and 25\textdegree{} ($\mathrm{NA}=0.09$, 0.17, 0.26, 0.34 and 0.42, respectively). 
Devices with $S \geq 0.8$ (dashed line) are considered diffraction limited. {\bf c)} Compression factor as a function of the thickness of layer 2 of the device in Fig~\ref{Fig:R=340}.
{\bf d)}~$\mathcal{R}/\mathcal{R}_\mathrm{target}$ as a function of the  targeted compression factor $\mathcal{R}_\mathrm{target}$ for devices formed of 9, 11, ..., 27, 29 layers for $\mathrm{NA}=0.17$. Increasing the number of layers enables devices with higher compression factors. This improvement plateaus for devices with more than 17 layers.}
\label{Fig:Sweep_results}
\end{figure}

The results from this study are summarized in Fig~\ref{Fig:Sweep_results}. 
Figure~\ref{Fig:Sweep_results}a shows the maximum  $\mathcal{R}$ obtained using the optimization procedure outlined above as a function of the device's NA independent of the number of layers. 
As mentionned above, the largest compression factor, $\mathcal{R}=340$, is obtained for devices with the smallest NA. 
Conversely, for the largest NA ($\mathrm{NA}=0.43$, corresponding to an acceptance half-angle of 25\textdegree{}), we obtain $\mathcal{R}=5.5$ in a \SI{1.61}{\micron}-thick device.

Figure~\ref{Fig:Sweep_results}b contains a summary of the performance of multilayer stacks comprising only 13 layers as a function of $\mathcal{R}_\mathrm{target}$ for a range of NAs. 
We observe a clear tradeoff between $S$ and $\mathcal{R}$ of a spaceplate for a given NA. 
Conversely, for a given $\mathcal{R}$, a higher NA results in lower device Strehl ratio. 
Though we do not reproduce it here, this trend holds for all optimized devices for every number of layers we explored (9, 11, ..., 27, 29 layers). 

In Fig.~\ref{Fig:Sweep_results}d, we explore the impact of the total number of layers on the performance of the device. 
Here, we fix the NA to 0.17, and optimize devices for a range of  $\mathcal{R}_\mathrm{target}$ and for a number of layers ranging from 9 to 29. 
Increasing the number of layers enables devices with better performance and higher compression factors, which makes intuitive sense as doing so expands the space of possible devices. 
Though it is not shown here, we note that this increase does not always substantially improve the Strehl ratio, which we attribute to challenges with retrieving the global maximum within the growing parameter spaces.

In Fig.~\ref{Fig:Sweep_results}c, we show that for a spaceplate with a given compression factor $\mathcal{R}$, when increasing the thickness of any one layer, the compression factor is roughly periodic. 
If the thickness of the layer is increased enough, the compression factor will return to approximately the same value it had before the layer thickness increase. 
This means there is wide latitude to make the spaceplate thicker without substantially decreasing its performance. 
The periodicity suggests that the optimization is non-convex and so it is not surprising that the global maximum was not always found.

Our search yielded a few other qualitative observations of interest. 
First, our optimization procedure favoured devices with an odd number of layers, and where the outermost layers had a higher refractive index.
Additionally, devices optimized for operation with p-polarized light tended to yield higher compression factors. 

\begin{figure}[htb]
\begin{centering}
\includegraphics[width=1\linewidth]{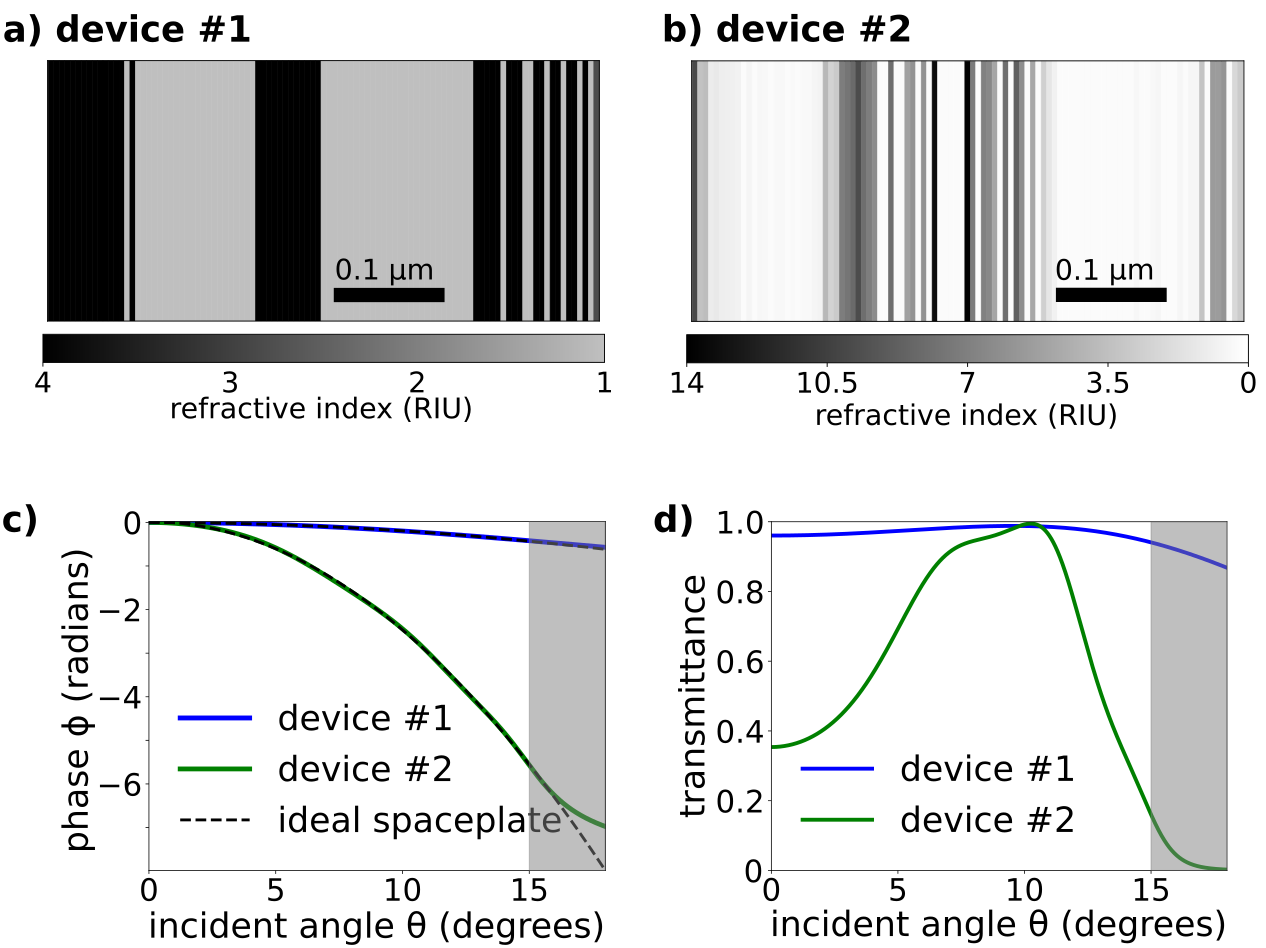}
\par\end{centering}
\caption{{\bf a)}~A spaceplate device with $\mathcal{R}=6.4$ designed by optimizing the refractive index of thin layers. Here, the index is bound within $1<n<4$. {\bf b)}~A spaceplate with $\mathcal{R}=80$, where $n$ is bound to positive values during optimization. Here, the resulting refractive index ranges 0.2 to 14. {\bf c)}~The transmission phases of the devices in (a) and (b) as a function of incident angle, and their respective fits to ideal spaceplate profiles. {\bf d)}~The normalized transmittance of the devices.}
\label{Fig:nonrealisticIndices}
\end{figure}

In our second approach to optimizing the spaceplate structures, we fix the total number of layers and their respective thicknesses and use the refractive indices of the layers as the optimization parameters. 
This approach is similar to the one explored in previous work featuring a different non-local operation in a thin-film multilayer stack~\cite{Silva2014}. 
Here, we set the thickness of 101 individual layers to 5~nm for a total device thickness of 505~nm, and we confine the refractive index to a realistic range for dielectric materials at optical frequencies ($n=1$~--~$4$). 
Our search yielded a device with a maximum compression factor of $\mathcal{R}=6.4$ (Fig.~\ref{Fig:nonrealisticIndices}a).
We also then repeat this procedure with the same parameters but expanding the range of allowed refractive indices to include any positive value. 
This range explicitly includes materials for which $n<1$, so-called epsilon-near-zero media~\cite{Alu2007, Reshef2019}. 
Though this optimization yields less feasible devices, it may provide information on the ultimate performance of spaceplates. 
Here, our search yielded a device with a maximum compression factor of  $\mathcal{R}=80$ for $\mathrm{NA}=0.26$, with refractive indices ranging from $n=0.2$ to $n=14$ (Fig.~\ref{Fig:nonrealisticIndices}).

Our study of the first approach suggests that the maximum compression factor is limited by the NA of the device. 
For example, for devices with an identical $S$ and number of layers, reducing the NA from 0.26 to 0.17 had the effect of nearly doubling $\mathcal{R}$ from 9.6 to 18.0.
This observation echoes a recent finding that the total effective propagation distance $d_\mathrm{eff}$ of resonator-based spaceplates is limited by the NA~\cite{Chen2020}. 
Despite this limit, we find that spaceplates based on this platform are shown to enable diffraction-limited operation with record high compression factors for large numerical apertures.

We had expected the second approach to yield devices with larger compression factors since it considered a larger parameter space with access to many more degrees of freedom. 
However, the best compression factor obtained by this approach, $\mathcal{R}=6.4$ fell short of  the solution found in the previous search for $\mathrm{NA}=0.26$, which yielded a maximum compression factor of $\mathcal{R}=11.5$. 
We suspect that the dramatic increase in optimization parameters hampered locating the global maximum for this class of devices. 
Nevertheless, increasing the range of allowable indices further increased the compression factor from $\mathcal{R}=6.4$ to 80, suggesting that $\mathcal{R}$ may also be limited by index contrast.

In summary, we have successfully inversely designed high-performing spaceplates based on multilayer thin-film stacks. 
We observed tradeoffs between the number of layers, NA, compression factor $\mathcal{R}$, and the degree of aberration (i.e., the Strehl ratio). 
While this work was focused on monochromatic spaceplates, these tradeoffs suggest there may be similar tradeoffs with the spaceplate's spectral bandwidth. 
Broadband operation and achromatic design are topics for future research. 
We observe signs that our gradient ascent is non-convex which suggests that other inverse-design schemes might be even more fruitful than gradient descent and, thus, may fully capitalize on the enormous design parameter space of the thin-film platform. 
Finally, while previous spaceplate designs have achieved very high $\mathcal{R}$~\cite{Guo2020a}, this work demonstrates that even higher $\mathcal{R}$ can be achieved in thin-film stacks, a mature commercial fabrication platform.

\section*{Methods}
Our optimization starts by initializing a group of 200 spaceplate candidate designs, each with randomly generated layer thicknesses or refractive indices. We pre-define the number of layers, the targeted compression factor  $\mathcal{R}$, and the range of incident angles of light (i.e., the NA). 
The code then applies the gradient descent algorithm to each layer of the spaceplate, differentiating by either the layer thickness or the refractive index to find the respective gradient. 
For a given candidate design, the algorithm starts by simultaneously optimising the first 3 layers, and once they are fully optimised, it includes the next layer into the simultaneous optimization process.
The spaceplate performance is assessed once all of the layers have been sequentially included and optimized. 

The FOM is the inverse RMSE $\sigma$, as calculated by 

\begin{equation}
    \sigma^2 = \frac{1}{N}\sum_{i=0}^{N}(\phi(\theta_{i})_{\mathrm{target}}-\phi(\theta_{i}))^2,
\end{equation}\\ 
where $N$ is the number of points between $\theta_0$ and $\theta_{\mathrm{max}}$, $\phi(\theta_i)$ is obtained from TMM for the designed spaceplate and $\phi(\theta_{i})_\mathrm{target}$ is the phase of an ideal spaceplate with $d_\mathrm{eff}=\mathcal{R}_{\mathrm{target}}d$. The incident angles range from $\theta_0=0$\textdegree{} to $\theta_\mathrm{max}=\sin^{-1}{(\mathrm{NA})}$.
This FOM was maximized in the first optimisation approach which focused on layer thicknesses. In the second approach, which focused on refractive index, the FOM was changed to $1/(\sigma (1 - \langle 10\log_{10}(T) \rangle))$ where $\langle 10\log_{10}(T) \rangle$ is the insertion loss in dB averaged over angle across the NA of the device. In this approach, the total number of layers is substantially larger than in the devices explored previously. This optimization approach is much more numerically cumbersome and so we are forced to iterate over fewer parameters. 
Consequently we optimize a single candidate device at a time rather than a swarm of 200.

\section*{Author contributions}
OR and JSL conceived the basic idea for this work. JTRP and OR performed the numerical calculations. JTRP and OR analysed the numerical results. RWB and JSL supervised the research and the development of the manuscript. JTRP wrote the first draft of the manuscript, and all authors subsequently took part in the revision process and approved the final copy of the manuscript. 
Portions of this work were presented at Frontiers Washington, DC in 2019~\cite{Page2020}.

\section*{Acknowledgements}
The authors thank Marc-Andr\'{e} Renaud for programming advice, and V\'{i}ctor J. L\'{o}pez-Pastor and Ali H. Alhulaymi for helpful discussions.

The authors acknowledge support from the Transformative Quantum Technologies program of the Canada First Research Excellence Fund, the Canada Research Chairs Program, and the Natural Sciences and Engineering Research Council of Canada.

\centering
\onecolumn
For Table of Contents Only
\begin{figure}[H]
\begin{centering}
\includegraphics[width=3.25in, height=1.75in]{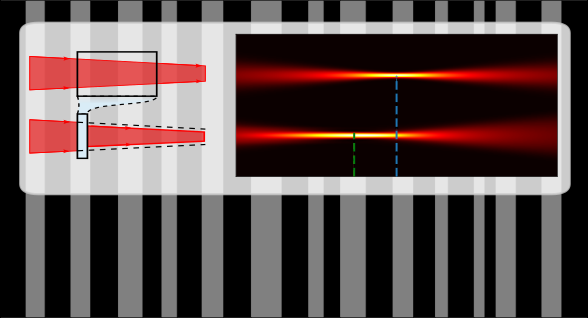}
\par\end{centering}
\end{figure}

\end{document}